# Influence of Se doping in recently synthesized NaInS$_{2-x}$Se$_x$ solid solutions for potential thermo-mechanical applications studied via first-principles method

M. M. Hossain[1,*], M. A. Ali[1], M. M. Uddin[1], M. A. Hossain[2], M. Rasadujjaman[3], S. H. Naqib[4,**], M. Nagao[5], S. Watauchi[5] and I. Tanaka[5]

[1]Department of Physics, Chittagong University of Engineering and Technology, Chattogram-4349, Bangladesh
[2]Department of Physics, Mawlana Bhashani Science and Technology University, Santosh, Tangail-1902, Bangladesh

[3]Department of Physics, Dhaka University of Engineering & Technology (DUET), Gazipur, Gazipur-1700, Bangladesh

[4]Department of Physics, University of Rajshahi, Rajshahi 6205, Bangladesh
[5]Center for Crystal Science and Technology, University of Yamanashi, 7-32 Miyamae, Kofu, Yamanashi 400-8511, Japan

Corresponding author, *email: mukter_phy @cuet.ac.bd; mukter.phy@gmail.com
**email: salehnaqib@yahoo.com

Abstract

In the present work, the structural and hitherto unexplored thermal and mechanical properties of NaInS$_{2-x}$Se$_x$ (x = 0, 0.5, 1.0, 1.5 and 2.0) compounds have been studied using the density functional theory. Besides, the elastic anisotropy indices and hardnesses of NaInS$_{2-x}$Se$_x$ have been investigated as Se content is varied. The mechanical stability of all the compounds under study has been confirmed. The ratio of shear to bulk modulus ($G/B$) is low suggesting that the NaInS$_{2-x}$Se$_x$ (x = 0.5 and 1.5) compounds exhibit damage tolerant (ductility) properties while rest of the compositions are brittle in nature. The predicted hardness ($H$) values are also influenced with the Se content in the following order: $H$ (NaInSSe) > $H$ (NaInS$_2$) > $H$ (NaInSe$_2$) > $H$ (NaInS$_{1.5}$Se$_{0.5}$) > $H$ (NaInS$_{0.5}$Se$_{1.5}$). All the anisotropic indices under study indicate that NaInS$_{2-x}$Se$_x$ compounds are anisotropic in nature. The Mulliken bond population analysis suggests that the degree of covalency of In-S/Se bonds decreases when S is substituted by Se. The origin of low Debye temperature ($\Theta_D$) and low minimum thermal conductivity ($K_{min}$) have been successfully explained by considering the mean atomic weight ($M/n$) and average bond strength of the compounds. Temperature dependence of heat capacities ($C_v$, $C_p$) and linear thermal expansion coefficient ($\alpha$) are also estimated using the quasi-harmonic Debye model and discussed. The low values of $K_{min}$, $\Theta_D$ and $\alpha$ and damage tolerant behavior clearly indicate that the NaInS$_{2-x}$Se$_x$ (x = 0.5 and 1.5) compounds can be used as promising thermal barrier coating materials for high temperature applications.

1. Introduction

Chalcogenide compounds have the general formula $ABX_2$, where A is an alkali metal, B is a trivalent ion and X is O, S or Se, which follows α-$NaFeO_2$ layered type crystal structure with space group ($R\bar{3}m$) and point group ($D_{3d}^5$) [1, 2]. During the last few decades, chalcogenide materials have attained ample attention due to their fascinating structure and unique stoichiometry-controlled electro-optical properties [3-8]. Various physical properties of some chalcogenide compounds such as $CuInSe_2$, $CuInGaSe_2$, CdTe, and $Hg_{1-x}Cd_xTe$ having zinc blende or chalcopyrite structure have been studied. These materials are favourable to be used in photovoltaic devices and infrared detection applications [9-14]. In particular, chalcogenide semiconductor materials like $NaInS_2$ are promising photocatalyst for $H_2$ generation under the visible light irradiation [5]. The promising $NaInS_2$ nanocrystals and $NaInO_2$ materials were synthesized by different synthetic routes [1,6]. Electronic band structure and density of states for the $NaInO_2$ and $NaInS_2$ using a first-principles molecular orbital (MO) cluster calculations have also been reported [1] and the origin of the energy difference between filled valence band and empty conduction band as well as bonding nature have been discussed [1] . Physical properties of chalcogenide materials can be altered by alloying with different elements and by controlling particle size and crystal structure. Very recently, N. Takahashi *et al.* [3] have synthesized layered chalcogenides, $NaInS_{2-x}Se_x$ (x = 0, 0.5, 1.0, 1.5 and 2.0) by ball milling and subsequent heating. They studied crystal structure, electronic band structure and optical reflectance spectra experimentally and/or theoretically and suggested the possibility of $NaInS_{2-x}Se_x$ in the optoelectronic device applications under the visible light exposure. Apart from the well known optoelectronic applications of these compounds, search of novel possible applications in relation to desirable thermo-mechanical physical properties like low thermal conductivity, low density, low Debye temperature, high temperature stability and ability to protect mechanical damage could be quite interesting. It should be noted here that N. Takahashi *et al.* have synthesized the titled solid solutions in the temperature range 300 - 900 K which indicates the possible use of these materials at high temperatures [3].

It is instructive to note that some researchers have also studied various physical properties of similar types of compounds. It is reported that the structural, electronic and optical properties of $LiInSe_2$ and $LiInTe_2$ chalcopyrite have been studied via first-principles investigations [15]. This study [15] found that when Se is substituted by Te, the degree of covalency increased and also the refractive index. The electronic, optical, lattice dynamic and thermodynamic properties of $LiInX_2$ (X = S, Se, Te) were reported via first-principles studies [16, 17]. Temperature dependent thermodynamic properties with possible applications were also studied extensively. It was noticed that the structural stiffness was increased as one moved from LiInTe$_2$ to LiInS$_2$ and LiInSe$_2$ [16, 17]. Mechanical and optical properties of LiBX$_2$ (B = Ga, In; X = S, Se, Te) compounds were also studied [18]. Different elastic moduli, plasticity, and ductility behavior of these compounds were prominently reported. Y Ren *et al.* [19] reported the structural, mechanical properties, Debye temperature and minimum thermal conductivity of CuInS$_2$ compound. All these prior investigations demonstrate the promise of ternary chalcogenide compounds in various optoelectronic and thermo-mechanical applications.

Many physical properties appear to depend sensitively on substitutional composition with appropriate atomic species, which is of keen interest to the materials science researchers. To the best of our knowledge, no report exists on mechanical and thermodynamic properties of NaInS$_{2-x}$Se$_x$ (x = 0, 0.5, 1.0, 1.5 and 2.0) chalcogenide compounds. Only synthesis process, electronic properties and reflectivity spectra were reported [3]. During designing a new device such as photovoltaic or optoelectronic device, knowledge of mechanical properties of the active material is essential. Moreover, temperature and pressure dependent thermodynamic properties enhance our understanding of the overall behavior of the solid of interest under different external conditions. It is expected that clear understanding and correlation between mechanical and thermodynamic properties of a material can help greatly to design materials for device applications. All these considerations motivate us to perform a first principles study on NaInS$_{2-x}$Se$_x$ solid solutions. In this work, we successfully investigated the effects of Se substitution for S atom in the NaInS$_2$ on the structural, mechanical and thermodynamic properties of NaInS$_{2-x}$Se$_x$ (x = 0.5, 1.0, 1.5, 2.0) solid solutions for the first time. We also investigated their mechanical stability, hardness and elastically anisotropic behaviour.

2. Computational Methodology

The high-throughput density functional theory (DFT) calculations for NaInS$_{2-x}$Se$_x$ (x = 0, 0.5, 1.0, 1.5 and 2.0) chalcogenide compounds are done using the CAmbridge Serial Total Energy Package (CASTEP) Code [20]. The ground state energy of the system is found by solving the Kohn-Sham equation within the generalized gradient approximation (GGA) for electron exchange-correlations, where the functional form is of Perdew-Burke-Ernzerhof (PBE) type [21, 22]. Broyden Fletcher Goldfarb Shanno (BFGS) geometry optimization method is employed to optimize the atomic site configuration [23]. The crystal structures are optimized using the ultra-soft pseudopotential [24, 25] with cut-off energy of 450 eV for the plane wave expansion and k-point sampling at 9×9×2 special points in a Monkhorst-Pack grid [26] for all the compositions. The convergence thresholds were as follows: energy tolerance was 5 × 10$^{-6}$ eV atom$^{-1}$, maximum force was 0.01 eV Å$^{-1}$, maximum stress was 0.02 GPa and maximum displacement was 5 × 10$^{-4}$ Å for geometry optimization. The 'stress-strain' method contained within the CASTEP program was employed to calculate the single crystal elastic constants. Virtual Crystal Approximation (VCA) approach for NaInS$_{2-x}$Se$_x$ (x = 0.5, 1.0, 1.5, 2.0) solid solutions was implemented in the study of mechanical properties within the context of density functional methods followed by the *stress-strain* method [27]. This technique allows one to mix different atoms, isotopes and oxidation states for a specific atomic site to be occupied randomly. VCA does not allow any probable short-range order. It is assumed that a virtual atom occupies every potentially disordered site and extrapolates an averaged behavior between the real components.

3. Results and Discussion

3.1 Structural properties

The chalcogenide compound, $NaInS_2$ and its solid solution compositions, as mentioned previously, have an α-$NaFeO_2$ type crystal structure [1, 2]. The crystal structure of $NaInS_2$ as the prototype structure is shown in **Fig. 1**. The Wyckoff atomic positions are as follows: Na (0, 0, 0.5), In (0, 0, 0), S/Se (0, 0, 0.26). The crystallographic lattice constants at ambient condition for pure and Se containing $NaInS_{2-x}Se_x$ compounds were obtained after optimizing the crystal structure and are shown in **Table 1**, together with previously reported available values for comparison.

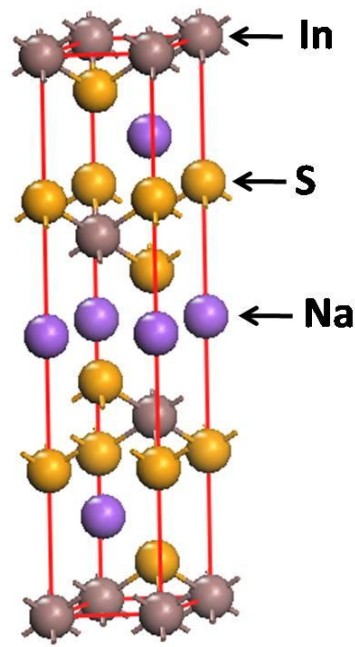

**Fig. 1:** The crystal structure of $NaInS_2$ conventional unit cell.

The lattice constants are found to increase with increasing Se substitution in the $NaInS_{2-x}Se_x$ as depicted in **Fig. 2**. The possible reason might be the difference between ionic radii of S (0.43Å) and Se (0.56 Å). The well known Vegard's law [28] is followed by the compositions under study. This largely validates the VCA approach adopted in this study. The deviation of the present calculations from the reported experimental work is less than 1.7 %. However, very similar trends for both experimental and theoretical results are noticed (**Fig. 2**) and it can be concluded that the obtained lattice constants are reasonable in comparison with prior experimental data. This confirmed the reliability of our calculations presented in this paper.

**Table 1:** Optimized lattice parameters, *a*, *b* and *c* (all in Å), unit cell volume *V* (Å$^3$) of NaInS$_{2-x}$Se$_x$ (x = 0, 0.5, 1.0, 1.5 and 2.0) solid solutions.

| Optimized parameters | a = b | % of deviation | c | V | % of deviation | Ref. |
|---|---|---|---|---|---|---|
| NaInS$_2$ | 3.868 | 1.70 | 20.134 | 260.887 | 1.22 | This |
|  | 3.803 |  | 19.89 |  |  | Expt.[3] |
| NaInS$_{1.5}$Se$_{0.5}$ | 3.906 | 1.69 | 20.379 | 269.275 | 1.08 | This |
|  | 3.841 |  | 20.160 |  |  | Expt.[3] |
| NaInSSe | 3.935 | 1.33 | 20.501 | 274.934 | 0.47 | This |
|  | 3.883 |  | 20.405 |  |  | Expt.[3] |
| NaInS$_{0.5}$Se$_{1.5}$ | 3.973 | 1.32 | 20.779 | 284.113 | 0.80 | This |
|  | 3.921 |  | 20.614 |  |  | Expt.[3] |
| NaInSe$_2$ | 4.022 | 1.25 | 21.005 | 294.279 | 0.55 | This |
|  | 3.972 |  | 20.890 |  |  | Expt.[3] |

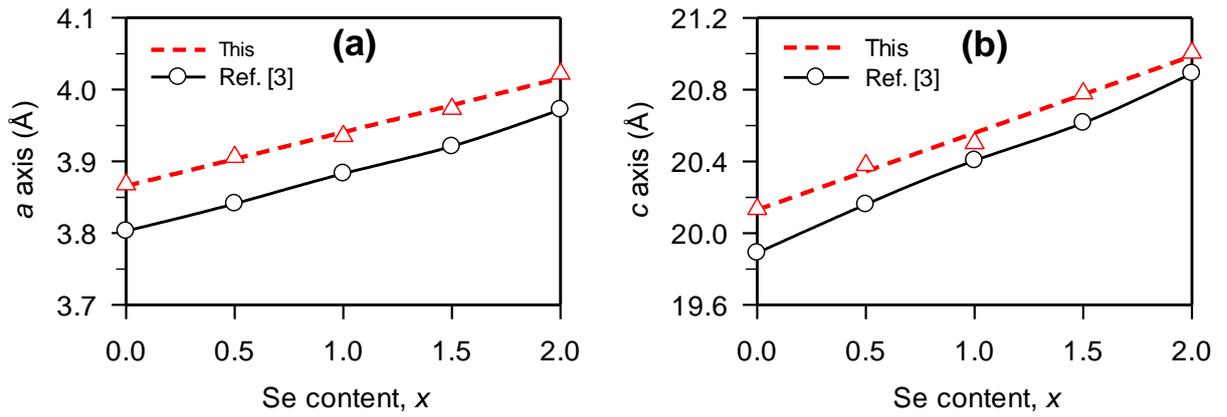

**Fig. 2:** Variation of calculated lattice parameters and experimental data of Ref. [3] (a) *a* and (b) *c* as a function of Se content, *x* in NaInS$_{2-x}$Se$_x$ (x = 0, 0.5, 1.0, 1.5 and 2.0) solid solutions. The dashed straight lines represent the best fits to the computed values.

3.2 Elastic constants, mechanical stability and Cauchy pressure

The interest in the study of elastic constants, at present is significant because the elastic constants can link the stress to strain behavior and describe the response of crystals under external forces of various types. Furthermore, this can estimate the technologically important mechanical properties such as stiffness, hardness, brittleness/ductility, machinability index, elastic anisotropy, different types of elastic moduli, Poisson's ratio, Cauchy pressure indicator etc. Different thermodynamic parameters such as Debye temperature, minimum thermal conductivity, specific heat, melting point etc. are also closely related to elastic parameters and can be estimated depending on their specific connection with elastic stiffness constants.

F. Mouhat *et al.* [29] has described the conditions of mechanical stability of unstressed crystalline structures for all crystal symmetries, while the Born elastic stability criteria are most suitable for cubic crystals [30]. Since all the compounds under study are hexagonal, they should have five independent elastic constants. We have summarized all these constants in **Table 2**, with reported available data for comparison.

**Table 2:** The calculated single crystal elastic constants $C_{ij}$ (GPa), Cauchy pressure ($CP_x$, $CP_y$) (GPa) and Poisson's ratio ($\upsilon$) for NaInS$_{2-x}$Se$_x$ (x = 0, 0.5, 1.0, 1.5 and 2.0) solid solutions.

| Compounds | $C_{11}$ | $C_{33}$ | $C_{44}$ | $C_{12}$ | $C_{13}$ | $C_{66}$ | $CP_x$ | $CP_y$ | $\upsilon$ | Ref. |
|---|---|---|---|---|---|---|---|---|---|---|
| NaInS$_2$ | 99 | 70 | 24 | 33 | 22 | 33 | -2 | 0 | 0.246 | This |
| LiInS$_2$ | 50.1 | 61.7 | 20.8 | -11.7 | -3.5 | 36.4 | - | - | - | [31] |
| LiInS$_2$ | 70.8 | 66 | 18.3 | 45.8 | 40.4 | 14.2 | - | - | - | [32] |
| NaInS$_{1.5}$Se$_{0.5}$ | 92 | 65 | 15 | 30 | 19 | 31 | 4 | -1 | 0.277 | This |
| NaInSSe | 95 | 75 | 28 | 33 | 25 | 31 | -3 | 2 | 0.244 | This |
| NaInS$_{0.5}$Se$_{1.5}$ | 87 | 81 | 17 | 33 | 26 | 27 | 9 | 6 | 0.297 | This |
| NaInSe$_2$ | 85 | 66 | 25 | 29 | 23 | 28 | -2 | 1 | 0.243 | This |
| LiInSe$_2$ | 50.3 | 58.2 | 12.6 | 25.8 | 19.6 | 13.6 | - | - | - | [32] |

It is found from **Table 2** that all elastic constants are non zero and positive, and obey the necessary and sufficient conditions of stability for the hexagonal crystal class. The complete conditions for mechanical stability are as follows [29]: $C_{11} > |C_{12}|$, $2C_{13}^2 < C_{33}(C_{11} + C_{12})$, $C_{44} > 0$. In this study, all compositions fulfilled the above mechanical stability criteria and hence, the NaInS$_{2-x}$Se$_x$(x = 0, 0.5, 1.0, 1.5, 2.0) materials are predicted to be mechanically stable.

The elastic stiffness constants data of the NaInS$_{2-x}$Se$_x$ solid solutions are compared with those of Li-containing chalcogenide LiInS$_2$ and LiInSe$_2$ as there is no available experimental data on NaInS$_{2-x}$Se$_x$ solid solutions. The titled compounds exhibit much higher values of elastic stiffness constants compared to the other two reported compounds (see **Table 2**) [31], [32]. Therefore, it is expected that the compounds under study could show strong elastic behaviour. However, three diagonal ($C_{11}$, $C_{33}$ and $C_{44}$) and two off-diagonal ($C_{12}$ and $C_{13}$) elastic constants are found in hexagonal system. The constant $C_{66}$ is not independent and can be obtained as ($C_{11}$-$C_{12}$)/2 and indicates the shear resistance along [110] direction (also shown in **Table 2**). All single crystal elastic constants, $C_{ij}$ are shown in **Fig. 3**. The elastic constants $C_{11}$ and $C_{33}$ represent the stiffness of a hexagonal crystal to unidirectional stress applied along [100] and [001] directions, respectively, whereas the $C_{44}$ and $C_{66}$ measure the resistance against shear deformation in the (100) plane. Some interesting mechanical features are noted below.

(i) The materials are more resistant to deformation due to external stress along [100] directions on (100) plane than that along [001]/[010] directions, as $C_{11} > C_{33}$. This is indicative of anisotropic bonding feature present in the compounds. It means that the bonding strength along *a* axis is stronger than that along the *c* axis.

(ii) The values of $C_{11}$ and $C_{33}$ are much higher than that of $C_{44}$. This indicates that the linear elastic deformation is much harder along [100] and [001] directions than the shear deformations in the (100) plane with a stress in the [010] direction.

(iii) It is apparent from **Table 2** that $C_{66} > C_{44}$ for all the considered compounds, which suggest that there is higher resistance to basal and prismatic shear deformations along [110] direction compared to that in the (100) plane along [010] direction.

(iv) The large differences between the $C_{12}$ and $C_{13}$ are observed for all compositions. This implies that all the titled compositions have greater ability to resist shear distortion under applied force along the *b* and *c* axes than along the *a* axis.

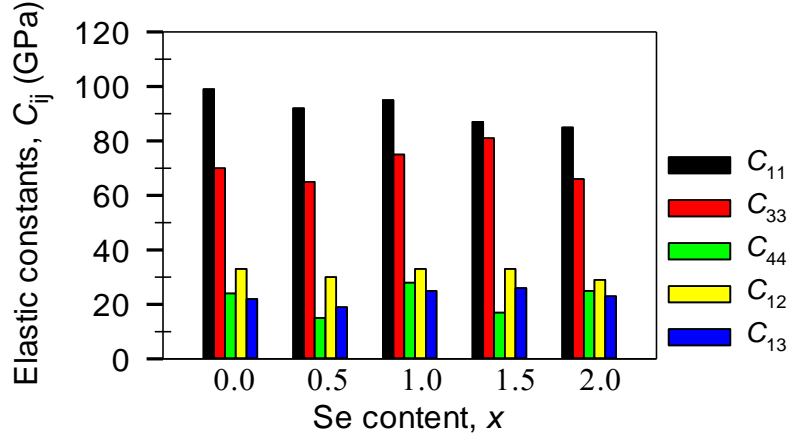

**Fig. 3:** Single crystal elastic constants, $C_{ij}$ (GPa) for NaInS$_{2-x}$Se$_x$ (x = 0, 0.5, 1.0, 1.5 and 2.0) solid solutions.

Cauchy pressure (CP) is a parameter that has been used to identify the type of chemical bonding present in a compound. Conventionally, the sign of CP predicts the nature of bonding and for hexagonal system; the CP is calculated for different directions by the following formulae [33]: $CP_x = C_{13} - C_{44}$ and $CP_y = C_{12} - C_{66}$. If a material has a negative (positive) CP, covalent (ionic) bonding will be dominant. As shown in **Table 2**, it is found that the compounds NaInS$_{1.5}$Se$_{0.5}$, NaInSSe and NaInSe$_2$ contain mixed nature (ionic and covalent) of bonding due to the different signs of $CP_x$ and $CP_y$, while the positive sign of both $CP_x$ and $CP_y$ indicates ionic character for the NaInS$_{0.5}$Se$_{1.5}$ composition. On the other hand, the sign of $CP_x$ is negative, indicating ionic nature but the value of $CP_y$ is zero for NaInS$_2$ compound, which suggests the atoms are interacting through a central force potential [34] in this solid. The type of chemical bonding can also be predicted using the value of Poisson's ratio ($v$). The relevant formula to calculate $v$ can be found elsewhere [35]. When the value of $v$ is greater (smaller) than 0.25, the chemical bonding is ionic (covalent) in nature [33, 34]. The values of $v$ for the NaInS$_{2-x}$Se$_x$ (x = 0, 0.5, 1.0, 1.5 and 2.0) are of 0.247, 0.277, 0.244, 0.297 and 0.243, respectively. These suggest that the ionic bonding dominates in the NaInS$_{1.5}$Se$_{0.5}$ and NaInS$_{0.5}$Se$_{1.5}$ compounds while NaInS$_2$, NaInSSe and NaInSe$_2$ compounds have mixed bonding character. It is interesting to note that prediction regarding bonding nature depending on Cauchy pressure and Poisson's ratio are largely consistent with each other.

### 3.3 Polycrystalline elastic moduli, hardness and machinability

The mechanical behaviour of solid materials under external stress is an important factor that needs to be considered when selecting the compounds for applications in various engineering fields. The polycrystalline elastic moduli such as bulk modulus (*B*), shear modulus (*G*) and Young modulus (*Y*) as well as ductile/brittle behaviour were calculated from single crystal elastic constants and are listed in **Table 3**, together with previously reported available Li-containing chalcogenide data for comparison only as there is no available experimental data on different elastic moduli of NaInS$_{2-x}$Se$_x$ solid solutions. The polycrystalline moduli $B$ and $G$ were estimated following Voigt-Reuss-Hill approximations [36-38] and subsequently Young modulus, $Y$ was calculated by the well established formula: $Y = 9BG/(3B+G)$ [38]. We have summarized all these values in **Table 3**.

**Table 3:** The calculated bulk modulus, $B$ (GPa), shear modulus, $G$ (GPa), Young's modulus, $Y$ (GPa), Pugh ratio $G/B$, machinability index ($B/C_{44}$), micro hardness, $H_{\text{micro}}$ (GPa), macro hardness, $H_{\text{macro}}$ (GPa) and Lame's constant ($\lambda, \mu$) (GPa) for NaInS$_{2-x}$Se$_x$ (x = 0, 0.5, 1.0, 1.5 and 2.0) solid solutions.

| Compounds | $B$ | $G$ | $Y$ | $G/B$ | $B/C_{44}$ | $H_{\text{micro}}$ | $H_{\text{macro}}$ | $\lambda$ | $\mu$ | Ref. |
|---|---|---|---|---|---|---|---|---|---|---|
| NaInS$_2$ | 46 | 28 | 70 | 0.61 | 1.92 | 4.72 | 4.86 | 27.20 | 28.09 | This |
| LiInS$_2$ | 50.6 | 14.5 | 39.6 | - | - | - | - | - | - | [32] |
| NaInS$_{1.5}$Se$_{0.5}$ | 41 | 22 | 56 | 0.52 | 2.80 | 3.27 | 2.73 | 27.23 | 21.93 | This |
| NaInSSe | 47 | 29 | 72 | 0.62 | 1.67 | 4.95 | 5.15 | 27.58 | 28.94 | This |
| NaInS$_{0.5}$Se$_{1.5}$ | 47 | 23 | 57 | 0.48 | 2.76 | 2.97 | 2.02 | 32.15 | 21.97 | This |
| NaInSe$_2$ | 42 | 26 | 65 | 0.62 | 1.68 | 4.45 | 4.67 | 24.72 | 26.15 | This |
| LiInSe$_2$ | 31.3 | 13.1 | 34.4 | - | - | - | - | - | - | [32] |

It is seen from table that all the Na-containing chalcogenides under study show much higher values of $B$, $G$, and $Y$ compared to the Li-containing chalcogenides except for $B$ of LiInS$_2$ [32]. Higher bond density, shorter bond length and greater degree of covalency are key factors for higher elastic moduli as well as hardness [40]. Thus, comparatively lower values of elastic moduli were found due to the higher bond length and much lower degree of covalency of Na-S/Se bonds, which can be clearly seen from Mulliken bond population analysis which will be discussed in a later section. However, as seen from **Table 3**, the highest value of three elastic moduli, $Y$ (72 GPa), $G$ (29 GPa) and $B$ (47 GPa) was found for the NaInSSe composition, while the NaInS$_{1.5}$Se$_{0.5}$ compound exhibits the lowest values of these three moduli. Therefore, among the studied compositions, materials could be reasonably stiff, resistant to shear deformation and volume deformation only with 50% replacement of S with Se in NaInS$_2$. On the other hand, on 25% replacement of S with Se in NaInS$_2$, the compound is predicted to show the lowest stiffness, the least resistance against reversible shape deformation and volume deformation. It is also observed that all the elastic moduli of end member, NaInS$_2$ is higher than that of the other end member, NaInSe$_2$. Therefore, it can be concluded that the polycrystalline elastic moduli are sensitive to Se substitution in the NaInS$_2$ compound and can be controlled with Se substitution. The ductility/brittleness can be predicted via the ratio of $G/B$, which was first proposed by Pugh [41]. A high value of $G/B$, greater than 0.57 of a material is associated with ductility and a lower value with brittleness. Therefore, considering the values of $G/B$ in **Table 3**, NaInS$_{1.5}$Se$_{0.5}$ and NaInS$_{0.5}$Se$_{1.5}$ compounds should be ductile in nature whereas rest of the materials (NaInS$_2$, NaInSSe and NaInSe$_2$) are expected to exhibit brittle behaviour. This is clearly depicted in **Fig. 4**. In addition to Pugh ratio, Poisson's ratio (υ) is also an important index to differentiate between brittle materials from the ductile ones. It is reported that a material behaves as brittle for a critical value of υ less than ~ 0.26 otherwise the ductile behavior of material is prominently noted [42]. Our calculated values once again confirm the ductile nature of NaInS$_{1.5}$Se$_{0.5}$ and NaInS$_{0.5}$Se$_{1.5}$ while the NaInS$_2$, NaInSSe and NaInSe$_2$ compounds are demonstrating brittle character. The results of elasticity can also be complemented using the Lame's constant (λ, μ) derived from $Y$ and $\upsilon$ [43] as shown in **Table 3**. The constant, λ measure the compressibility whereas μ measures the shear stiffness of a material. Our results suggest that the compound NaInSSe and NaInS$_{0.5}$Se$_{1.5}$ show comparatively higher level of shear stiffness and larger compressibility, respectively.

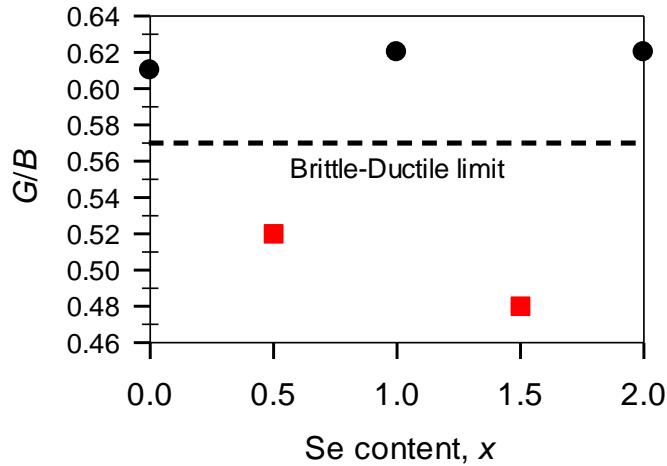

**Fig. 4:** $G/B$ ratio, called the Pugh ratio, distinguishing ductile materials from brittle materials for NaInS$_{2-x}$Se$_x$ (x = 0, 0.5, 1.0, 1.5 and 2.0) solid solutions as a function of Se content.

Hardness of a solid material is a useful parameter in the field of engineering to design different devices. It is possible to calculate the hardness value using the polycrystalline elastic moduli. Hardness of solids is related to the resistance against indentation. Here we have estimated two types of hardness; micro hardness ($H_{micro}$) and macro hardness ($H_{macro}$). $H_{micro}$ is linked to the Young modulus and Poisson's ratio, while $H_{macro}$, sometimes called the Chen hardness is estimated from shear and bulk moduli. Theoretical description underlying both the hardness formulae can be found elsewhere [44], [45]. The calculated hardnesses ($H_{micro}$ and $H_{macro}$) differ by less than 5%. Therefore, deviation of results for both calculations of hardness seems to be reasonable. It is noted that the values of $H_{micro}$ and $H_{macro}$ are in the range of 2.02 to 5.15 GPa, which suggest that all compounds under study are somewhat soft. The values of hardness for all compounds are comparable with the values (2 – 8 GPa ) of well known MAX phase and some other related compounds [46-48]. Moreover, the hardness values of all the studied compounds follow almost similar trend as the elastic moduli behavior with Se contents. To be specific, the NaInSSe (NaInS$_{2-x}$Se$_x$ with $x$ = 1.0) compound possesses the highest hardness value and can be considered as a reasonably hard material compared to other material compositions under study. Z. Sun *et al.* [49] defined machinability index ($B/C_{44}$) as the ratio of bulk modulus, $B$ and shear elastic constant, $C_{44}$. **Fig. 5** shows the variation of $C_{44}$ and $B/C_{44}$ with Se content in NaInS$_{2-x}$Se$_x$ solid solutions. The order of the $B/C_{44}$ is as follows: NaInS$_{1.5}$Se$_{0.5}$ > NaInS$_{0.5}$Se$_{1.5}$ > NaInS$_2$ > NaInSe$_2$ > NaInSSe. The variation of $B$ (ranging from 41 to 47 GPa) is much lower than that of $C_{44}$ (ranging from 17 to 28 GPa) with the increase of Se content. This indicates that the changes in machinability indices are mainly dependent on the shear stiffness constant, $C_{44}$ which is comparatively more sensitive with Se content of the studied compounds.

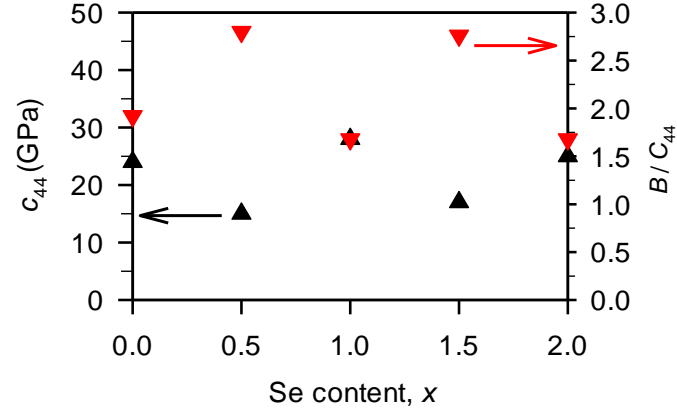

**Fig. 5:** $C_{44}$ (GPa) and the machinability index of NaInS$_{2-x}$Se$_x$ (x = 0, 0.5, 1.0, 1.5 and 2.0) solid solutions.

3.4 Elastic anisotropy

Elastic or mechanical anisotropy of materials is a quantitative measurement of direction dependent properties of systems under external stress. It is well known that most of the crystalline solid materials show anisotropy. Many physical processes such as the growth of plastic deformations in solids, development of crack and subsequently its propagation, microscale cracking in ceramics, fluid transport in poroelastic materials, internal friction, phase transformations etc. are significantly affected by the elastic anisotropy of the crystalline materials [50]. Therefore, the study of anisotropic behavior of solids is of notable interest from the point of view of practical applications.

**Table 4:** Shear anisotropic factors ($A_1$, $A_2$ and $A_3$), ratio of linear compressibilities along $c$ and $a$ axes ($k_c/k_a$), directional bulk modulus $B_a$ and $B_c$ (both in GPa) along $a$ and $c$ axes, respectively, percentage anisotropy factors in compressibility ($A_B$) and shear moduli ($A_G$), and universal anisotropic index, $A^U$ for NaInS$_{2-x}$Se$_x$ (x = 0, 0.5, 1.0, 1.5 and 2.0) solid solutions.

| Compound | $A_1$ | $A_2$ | $A_3$ | $k_c/k_a$ | $B_a$ | $B_c$ | $A_B$ | $A_G$ | $A^U$ |
|---|---|---|---|---|---|---|---|---|---|
| NaInS$_2$ | 1.27 | 0.72 | 2.78 | 1.83 | 139.42 | 145.76 | 2.06 | 1.63 | 0.20 |
| NaInS$_{1.5}$Se$_{0.5}$ | 1.95 | 0.48 | 2.83 | 1.82 | 128.33 | 128.33 | 2.17 | 7.93 | 0.90 |
| NaInSSe | 1.05 | 0.90 | 2.87 | 1.56 | 136.55 | 175.07 | 0.98 | 0.38 | 0.05 |
| NaInS$_{0.5}$Se$_{1.5}$ | 1.74 | 0.62 | 3.29 | 1.23 | 128.55 | 202.28 | 0.23 | 5.29 | 0.56 |
| NaInSe$_2$ | 1.02 | 0.89 | 2.75 | 1.58 | 121.85 | 159.49 | 1.03 | 0.42 | 0.06 |

Based on elastic constants and elastic moduli, many anisotropy indices under different conditions are considered to compute the overall degree of elastic anisotropy of NaInS$_{2-x}$Se$_x$. The relevant formulas for the materials having hexagonal symmetry are reported elsewhere [42, 46]. The results are summarized in **Table 4**. Before discussing the anisotropic behavior of solids, it is better to keep in mind that the values of most anisotropic factors yield unity when the crystalline material is isotropic, while a deviation (lesser or greater) from unity should refer to the degree of anisotropy. From **Table 4**, it is seen that the obtained values of shear anisotropic factors $A_1$, $A_2$ and $A_3$ for (100), (010) and (001) planes, respectively, are deviated from the unity, indicating anisotropic nature of all the compounds under study.

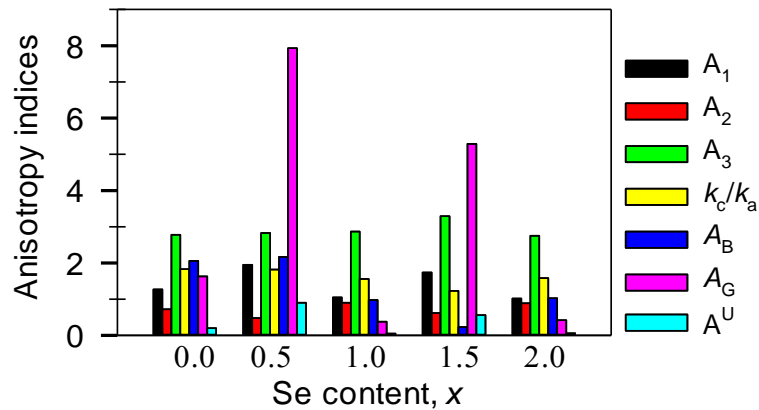

**Fig. 6:** Different anisotropy indices showing the degree of elastic anisotropy in NaInS$_{2-x}$Se$_x$ (x = 0, 0.5, 1.0, 1.5 and 2.0) compounds (See text for details).

Another anisotropy factor, called the linear compressibility anisotropy index is given by $k_c/k_a$, where $k_c$ and $k_a$ are the linear compressibilities along $c$ and $a$ axes which measure the degree of compressibility along these directions. This index also deviates from unity and suggest that the compressibility for the considered compounds along $c$ direction is higher than that along $a$ direction. Again, for the value of axial bulk modulus (in GPa) $B_c > B_a$ where $B_a$ and $B_c$ are measured along $a$ and $c$, respectively indicates that the bulk modulus along $c$ axis is much prominent than in $a$ direction for all compositions with the exception of NaInS$_{1.5}$Se$_{0.5}$. In other words, it is easier to compress the materials along $a$ direction than along the $c$ direction except for the compound of NaInS$_{1.5}$Se$_{0.5}$. In addition, it is also possible to estimate the percentage of elastic anisotropy for non-cubic polycrystalline solid systems. The percentage anisotropy in compressibility ($A_B$) and shear ($A_G$) can be calculated from bulk modulus (Voigt, Reuss) and shear modulus (Voigt, Reuss) from the following equations [50]: $A_B = \frac{B_V - B_R}{B_V + B_R} \times 100\%$ and $A_G = \frac{G_V - G_R}{G_V + G_R} \times 100\%$, where the subscripts, $V$ and $R$, indicate the Voigt and Reuss bounds, respectively. Ranganathan *et al.* [52] derived a useful parameter, $A^U$ called the universal anisotropic index applicable for all crystal symmetries. This anisotropy index can be defined as $A^U = 5\frac{G_V}{G_R} + \frac{B_V}{B_R} - 6 \geq 0$. The values of the factors $A_B, A_G$ and $A^U$ unambiguously suggest that all the compounds under study should exhibit elastically anisotropic behaviour under different modes of mechanical stress. All these elastic anisotropy indices can be easily visualized from **Fig. 6**.

For clear and complete understanding of the anisotropic behaviour of all the compositions under study, we have also made use of two dimensional (2D) and three dimensional (3D) plots. The details regarding these constructions can be found elsewhere [53]. In addition, all the elastic parameters in terms of minimum and maximum values are summarized in **Table 5**. Here, only the anisotropic 2D and 3D diagrams of NaInS$_2$ compound are shown in **Fig. 7** as the representative plot. When the plot yields spherical/circular shape, a perfectly isotropic behavior in 3D/2D is identified. The distortion from spherical shape indicates the degree of anisotropy in different directions in the 3D space.

**(a)** Young's modulus

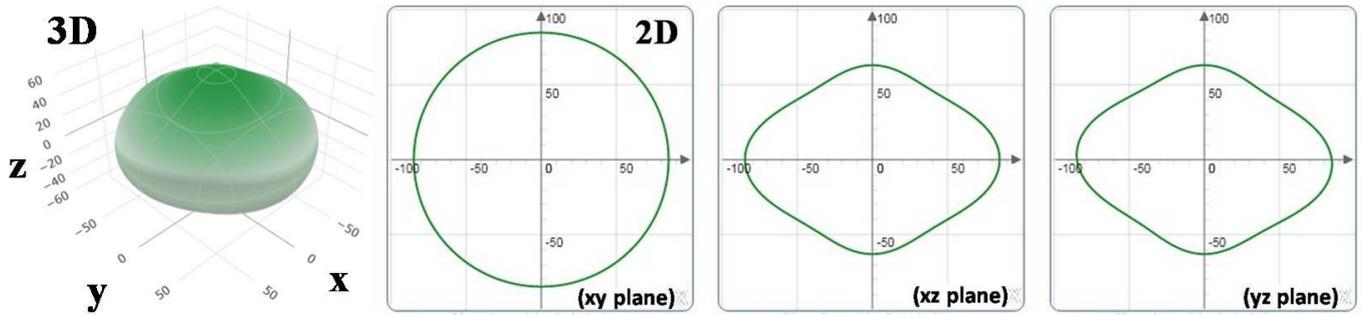

**(b)** Shear modulus

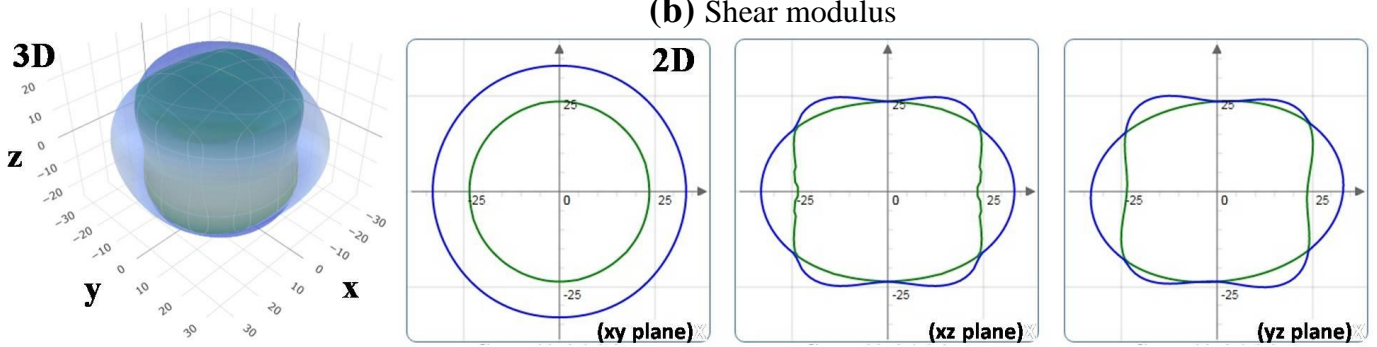

**(c)** Poisson's ratio

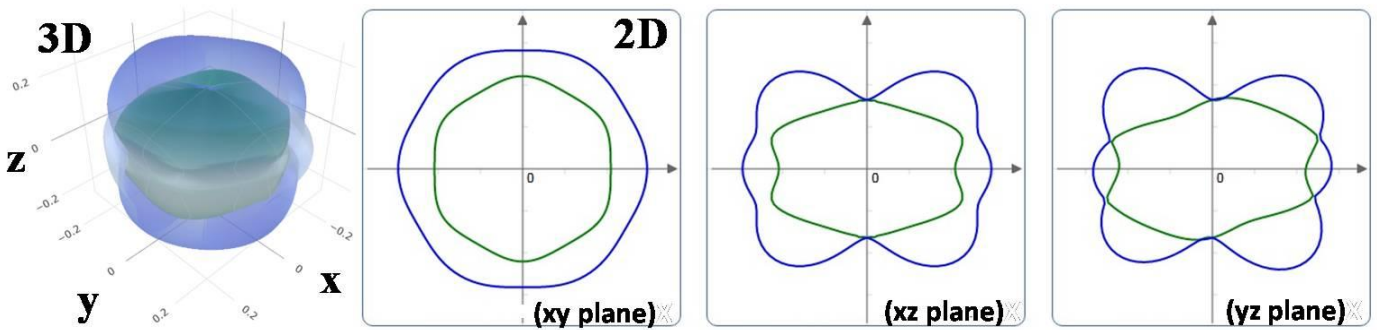

**Fig. 7:** The directional dependencies of three dimensional (3D) contour plots and two dimensional (2D) projections of (a) Young's modulus ($Y$) (b) shear modulus ($G$) and (c) Poisson's ratio ($\upsilon$) of $NaInS_2$.

**Table 5:** The limiting values (minimum and maximum) of Young's modulus, $Y$ (GPa), shear modulus, $G$ (GPa) and Poisson's ratio, $\upsilon$ for NaInS$_{2-x}$Se$_x$ (x = 0, 0.5, 1.0, 1.5 and 2.0) solid solutions.

| Compound | Young's modulus | | | Shear modulus | | | Poisson's ratio | | |
|---|---|---|---|---|---|---|---|---|---|
| | $Y_{min}$ | $Y_{max}$ | $Y_{max}/Y_{min}$ | $G_{min}$ | $G_{max}$ | $G_{max}/G_{min}$ | $\upsilon_{min}$ | $\upsilon_{max}$ | $\upsilon_{max}/\upsilon_{min}$ |
| NaInS$_2$ | 59.49 | 86.10 | 1.43 | 23.59 | 33.28 | 1.41 | 0.15 | 0.31 | 02.06 |
| NaInS$_{1.5}$Se$_{0.5}$ | 38.92 | 80.82 | 2.07 | 13.88 | 31.78 | 2.28 | 0.05 | 0.53 | 10.60 |
| NaInSSe | 64.59 | 79.78 | 1.23 | 27.35 | 31.31 | 1.14 | 0.19 | 0.28 | 01.47 |
| NaInS$_{0.5}$Se$_{1.5}$ | 41.25 | 74.50 | 1.80 | 15.07 | 29.48 | 1.95 | 0.05 | 0.54 | 10.42 |
| NaInSe$_2$ | 57.02 | 71.87 | 1.28 | 24.58 | 28.41 | 1.15 | 0.19 | 0.30 | 01.57 |

The Young's modulus, $Y$, shear modulus, $G$ and Poisson's ratio, $\upsilon$ show strongly anisotropic character in xz and yz planes while isotropic behaviour is found in the xy plane. Similar anisotropic behaviour is found in case of the other four compounds. Noted here that comparatively more pronounced anisotropic behaviour was found in case of two end members than in other compositions. Based on the anisotropy factor (AF) which in this case is defined by the ratio of maximum and minimum values of different elastic moduli, it is observed that the AF values of $Y$, $G$ and $\upsilon$ display the following sequence: $Y$, $G$, $\upsilon$ (NaInS$_{1.5}$Se$_{0.5}$) > $Y,G$, $\upsilon$ (NaInS$_{0.5}$Se$_{1.5}$) > $Y,G$, $\upsilon$ (NaInS$_2$) > $Y,G$, $\upsilon$ (NaInSe$_2$) > $Y,G$, $\upsilon$ (NaInSSe). This means that the highest anisotropy is found for the composition with 25% Se content while the lowest is for 50% Se content. It is interesting to note that the lowest anisotropy behaviour for the NaInSSe compound might be due to its highest value of elastic moduli as well as the hardness. The analysis presented in this paragraph agrees well with the analysis regarding various anisotropy indices in the preceding section. The behaviour of elastic anisotropy of various moduli under different external stress between two end members (NaInS$_2$ and NaInSe$_2$) increases in the order of $\upsilon$ - $Y$ - $G$ whereas for the three middle members (NaInS$_{2-x}$Se$_x$ with x = 0.5, 1.0 and 1.5) anisotropy increases in the sequence: $\upsilon$ - $G$ - $Y$.

### 4. Thermal properties

#### 4.1 Debye temperature

For many high temperature device applications, thermal barrier coating (TBC) materials are an important requisite. A popular application of TBC is the coating on turbine blades in gas-turbine engines which facilates its use at elevated operating temperature, resulting in the high efficiency of the engines. Therefore, to study the applicability of the NaInS$_{2-x}$Se$_x$ (x = 0, 0.5, 1.0, 1.5 and 2.0) materials as TBC, the phonon thermal conductivity ($K_{ph}$) as well as the minimum thermal conductivity ($K_{min}$) are needed to be considered. These two parameters rely on the Debye temperature ($\Theta_D$) and the Grüneisen parameter ($\gamma$). Debye temperature can be defined as the temperature at which all possible lattice vibrational modes in a crystal are excited. In other words, the Debye temperature corresponds to the maximum vibrational frequency, called Debye frequency in a crystal. Debye temperature, $\Theta_D$ can be calculated from the average sound velocity ($v_m$) which is correlated to shear modulus ($G$) and bulk modulus ($B$). The technologically important parameter $\Theta_D$ can be calculated as follows:

The mean sound velocity $v_m$ can be estimated from the following expression:

$$v_m = \left[1/3 \left(1/v_l^3 + 2/v_t^3\right)\right]^{-1/3}$$

where, $v_l$ and $v_t$ are longitudinal and transverse sound velocity, respectively. Again, $v_l$ and $v_t$ are related to the elastic moduli (shear and bulk modulus) and density of the solid as expressed by the following equations:

$v_l = [(3B + 4G)/3\rho]^{1/2}$ and $v_t = [G/\rho]^{1/2}$.

Finally, $\Theta_D$ can be calculated from:

$$\Theta_D = h/k_B \left[\left(3n/4\pi\right)N_A\rho/M\right]^{1/3} v_m,$$

where, $M$ denotes the molar mass, $n$ is the number of atoms in the molecule, $\rho$ is the mass density, and $h$, $k_B$, and $N_A$ are the Planck's constant, Boltzmann constant, and Avogadro's number, respectively. The calculated density ($\rho$), mean atomic weight ($M/n$), different sound velocities ($v_l$, $v_t$ and $v_m$) and $\Theta_D$ of all the compositions are tabulated in **Table 6** and also presented in **Fig. 8(a)**.

**Table 6:** Calculated crystal density, mean atomic weight ($M/n$) per atom, longitudinal, transverse and average sound velocities ($v_l$, $v_t$, and $v_m$), Debye temperature, $\Theta_D$, minimum thermal conductivity, $K_{min}$, lattice thermal conductivity, $k_{ph}$ at 300 K and Grüneisen parameter, $\gamma$ for NaInS$_{2-x}$Se$_x$ (x = 0, 0.5, 1.0, 1.5 and 2.0) solid solutions.

| Compounds | $\rho$ (kg/m³) | $M/n$ (amu/atom) | $v_l$(m/s) | $v_t$(m/s) | $v_m$(m/s) | $\Theta_D$ (K) | $K_{min}$ (W/mK) | $k_{ph}^*$(W /mK) | $\gamma$ | Ref. |
|---|---|---|---|---|---|---|---|---|---|---|
| NaInS$_2$ | 3856 | 50.49 | 4649 | 2695 | 2991 | 319 | 0.530 | 4.34 | 1.480 | This |
| LiInS$_2$ | - | - | 8871 | 4038 | 4552 | 306 | - | - | - | [32] |
| NaInS$_{1.5}$Se$_{0.5}$ | 4170 | 56.35 | 4136 | 2297 | 2558 | 270 | 0.443 | 2.08 | 1.638 | This |
| NaInSSe | 4509 | 62.21 | 4359 | 2536 | 2814 | 295 | 0.481 | 3.81 | 1.472 | This |
| NaInS$_{0.5}$Se$_{1.5}$ | 4774 | 68.07 | 3998 | 2147 | 2397 | 249 | 0.401 | 1.72 | 1.754 | This |
| NaInSe$_2$ | 5006 | 73.93 | 3913 | 2279 | 2528 | 259 | 0.413 | 3.62 | 1.467 | This |
| LiInSe$_2$ | - | - | 6720 | 3481 | 3897 | 244 | - | - | - | [32] |

The calculated $\Theta_D$ of the two end members under study is compared with those of reported Li-containing chalcogenides [32]. Anderson *et al.* [54] suggested that average atomic weight ($M/n$) strongly influences the $\Theta_D$. This indicates that higher the value of $M/n$, lower is the value of $\Theta_D$. Our results are roughly consistent with this suggestion. A low value of $\Theta_D$ signifies a low value of lattice thermal conductivity as well as a low minimum thermal conductivity. The highest $\Theta_D$ is found for NaInS$_2$ while it is the lowest in case of NaInS$_{0.5}$Se$_{1.5}$. The estimated values of $\Theta_D$ can be written in the following order: $\Theta_D$ (NaInS$_2$) > $\Theta_D$ (NaInSSe) > $\Theta_D$ (NaInS$_{1.5}$Se$_{0.5}$) > $\Theta_D$ (NaInSe$_2$) > $\Theta_D$ (NaInS$_{0.5}$Se$_{1.5}$) and average sound velocity of these compounds also follows the same order as the $\Theta_D$. The density and mean atomic weight increased systematically for NaInS$_{2-x}$Se$_x$ (x = 0, 0.5, 1.0, 1.5 and 2.0). All these imply together that the variation of $\Theta_D$ is primarily due to the variation of polycrystalline elastic moduli, especially the bulk and shear modulus.

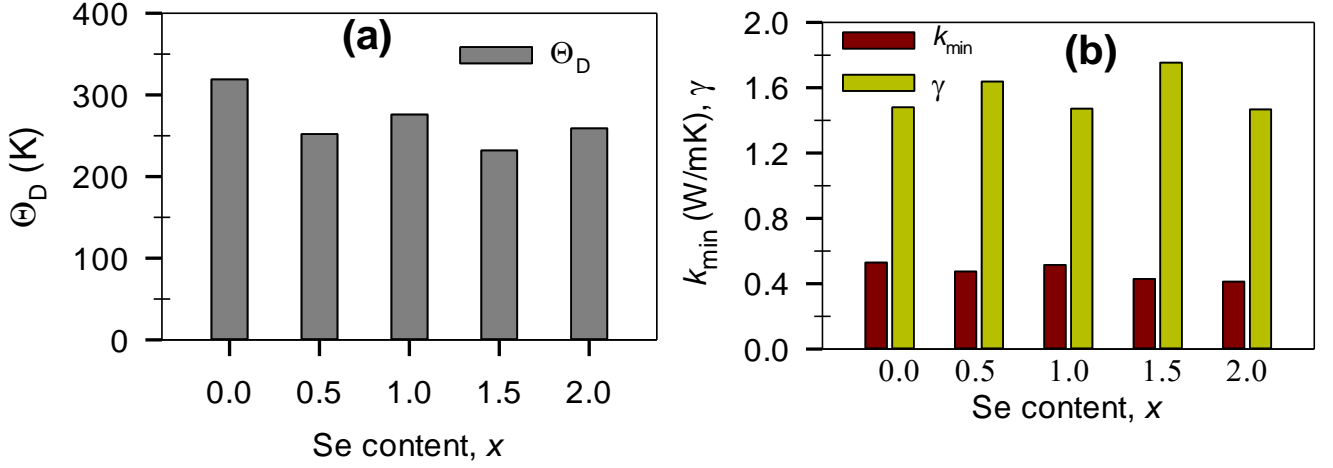

**Fig. 8:** Calculated (a) Debye temperature, $\Theta_D$ (K) and, (b) minimum thermal conductivity, $k_{min}$ (W/mK) and Grüneisen parameter ($\gamma$) for NaInS$_{2-x}$Se$_x$ (x = 0, 0.5, 1.0, 1.5 and 2.0) as a function of Se content.

4.2 Lattice thermal conductivity

The lattice thermal conductivity describes the heat conduction owing to the vibration of the lattice ions in a solid. To determine the lattice thermal conductivity ($k_{ph}$) of NaInS$_{2-x}$Se$_x$ (x = 0, 0.5, 1.0, 1.5 and 2.0) we have used the empirical formula derived by Slack [55] as follows:

$$K_{ph} = A(\gamma) \frac{M_{av} \Theta_D^3 \delta}{\gamma^2 n^{2/3} T}$$

$$\gamma = \frac{3(1+v)}{2(2-3v)}$$

where $\gamma$ is the Grüneisen parameter that measure the anharmonicity of phonons. In materials with high values of $\gamma$, significant anharmonic contributions are present and consequently such materials possesses low phonon thermal conductivity. In this study, we have found comparatively high value of $\gamma$ (see **Table 6 and Fig. 8 (b)**) for all compositions of NaInS$_{2-x}$Se$_x$ (x = 0, 0.5, 1.0, 1.5 and 2.0). Once the Grüneisen parameter is calculated, we can estimate the coefficient $A(\gamma)$ as follows.

$$A(\gamma) = \frac{4.85628 \times 10^7}{2(1 - \frac{0.514}{\gamma} + \frac{0.228}{\gamma^2})}$$

The theoretical lower limit of intrinsic thermal conductivity, which has been described by the modified Clarke's model [56] can be written as:

$$K_{min} = k_B v_m \left(\frac{M}{n\rho N_A}\right)^{-\frac{2}{3}}$$

The calculated minimum thermal conductivity, $K_{min}$, lattice thermal conductivity, $k_{ph}$ at temperature 300 K and the Grüneisen parameter, $\gamma$ are listed in **Table 6** and shown in **Fig. 8 (b).** The temperature dependent $k_{ph}$ for NaInS$_{2-x}$Se$_x$ (x = 0, 0.5, 1.0, 1.5 and 2.0) are depicted in **Fig. 9**. It should be pointed out that considering both Slack and modified Clarke model, $k_{ph}$ of NaInS$_{2-x}$Se$_x$ (x = 0, 0.5, 1.0, 1.5 and 2.0) decreased with temperature as 1303.7/T, 624.5/T, 1145/T, 518.9/T and 1086.1/T and subsequently $k_{ph}$ reaches almost constant value at high temperature.

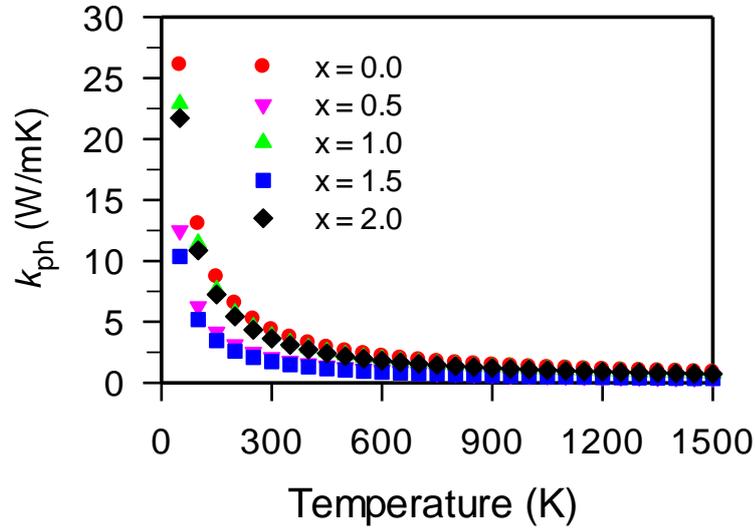

**Fig. 9:** Calculated lattice thermal conductivity, $k_{ph}$ (W/mK) for NaInS$_{2-x}$Se$_x$ (x = 0, 0.5, 1.0, 1.5 and 2.0) solid solutions as a function of temperature.

The origin of low $K_{min}$ for the compounds considered herein can be understood from the Mulliken bond overlap population analysis as shown in **Table 7**. It is noticed from the **Table 7** that the Mulliken populations of Na-S/Se bonds are smaller than that of In-S/Se bonds. These values indicate that covalent bonds exist in In-S/Se while Na-S/Se bonds show ionic nature. It has been reported [57] that strong ionic bonding and comparatively high Grüneisen parameters can lead to low thermal conductivity. Moreover, these softer Na-S/Se bonds with increase of temperature, reduce the phonon mean-free path to average atomic distance because the behavior of these bonds can be considered as equivalent to the thermal rattle structures, which provide weak zones that scatter phonons very efficiently [58]. Another possible reason, as suggested by Clarke, is that mean atomic weight ($M/n$) has the strongest influence on the $K_{min}$. Comparatively high values ranging from 50.49 to 73.94 amu/atom, are responsible for low thermal conductivity. The calculated $K_{min}$ and $\Theta_D$ values for NaInS$_{2-x}$Se$_x$ (x = 0, 0.5, 1.0, 1.5 and 2.0) are much smaller than that of a promising typical TBC material Y$_4$Al$_2$O$_9$ [58,59] and also compared to some other related compounds [19,51,60].

**Table 7:** Mulliken bond population and bond length of NaInS$_{2-x}$Se$_x$ (x = 0, 1.0 and 2.0) solid solutions.

| Compound | Bond | Bond number | Population | Length (Å) |
|---|---|---|---|---|
| NaInS$_2$ | S-In | 6 | 0.98 | 2.679 |
| | Na-S | 6 | 0.29 | 2.917 |
| | Na-S | 6 | 0.06 | 4.836 |
| NaInSSe | S-In | 3 | 1.08 | 2.711 |
| | Se-In | 3 | 0.40 | 2.757 |
| | Na-S | 3 | 0.33 | 2.916 |
| | Na-Se | 3 | 0.05 | 4.847 |
| NaInSe$_2$ | Se-In | 6 | 0.83 | 2.790 |
| | Na-Se | 6 | 0.34 | 5.047 |

4.3 Specific heat and thermal expansion coefficient

The specific heat at constant volume ($C_v$) can be well accounted for by the quasi-harmonic Debye model, at least for temperatures greater than 300 K (room temperature) as follows [61-64]:

$$C_v = 9nN_Ak_B\left(\frac{T}{\Theta_D}\right)\int_0^{x_D} dx \frac{x^4}{(e^x-1)^2} \text{ where, } x_D = \frac{\Theta_D}{T}$$

and $n$, $N_A$ and $k_B$ are the number of atoms per formula unit, Avogadro's number and Boltzmann constant, respectively. Again, the linear thermal expansion coefficient ($\alpha$), specific heat at constant pressure ($C_p$) are estimated according the following equations [61]:

$$\alpha = \frac{\gamma C_v}{3B_T v_m}$$

$$C_p = C_v(1 + \alpha\gamma T)$$

where, $B_T$, $v_m$ and $\gamma$ be the isothermal bulk modulus, molar volume and Grüneisen parameter, respectively.

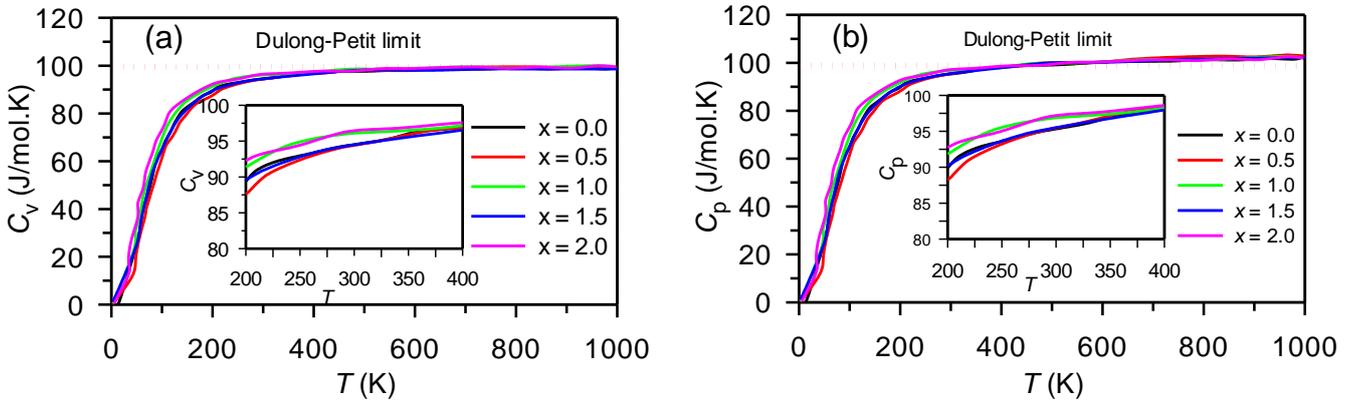

**Fig. 10:** Temperature dependent specific heat (a) $C_v$ at constant volume and (b) $C_p$ at constant pressure for NaInS$_{2-x}$Se$_x$ (x = 0, 0.5, 1.0, 1.5 and 2.0) solid solutions. The insets show magnified curves in the temperature range 200 K to 400 K.

The temperature dependences of specific heats, $C_v$ and $C_p$ in the temperature range 0 - 1000 K are shown in **Fig. 10.** The specific heats increase with the increasing temperature because of thermal softening. At low temperatures less than 300 K, the heat capacities obey the Debye $T^3$ power-law and at higher temperatures greater than 450 K, they almost follow the Dulong-Petit (DP) model. Although the $C_v$ at higher temperature (1000 K) approaches the DP limit ($3nN_Ak_B$), the DP model at higher temperature (1000 K), slightly underestimates the $C_p$, which is usually common for solid materials [63]. However, the deviations of $C_p$ values at temperature, 1000 K for all the titled compounds from the DP model were always less than 2.5%. The predicted value of $C_v$ has been compared with similar work reported by T. Ma *et al.*[16]. It is found that our calculated $C_v$ values of NaInS$_2$ and NaInSe$_2$ are 94.64 and 96.02 J/mol.K are higher than that (86.88 and 92.30 J/mol.K) reported for LiInS$_2$ and LiInSe$_2$ compounds, respectively.

We have also shown the temperature dependence of $\alpha$ in **Fig. 11**. It is found that $\alpha$ increased rapidly up to temperature 180 K and then gradually increased in the temperature ranging from 180 K to 300 K and finally followed an almost constant value. The values of $\alpha$ for NaInS$_{2-x}$Se$_x$ (x = 0, 0.5, 1.0, 1.5 and 2.0) solid solutions are $1.90 \times 10^{-5}$, $2.30 \times 10^{-5}$, $1.80 \times 10^{-5}$, $2.05 \times 10^{-5}$ and $1.70 \times 10^{-5}$ (all in K$^{-1}$) at temperature 300 K, respectively.

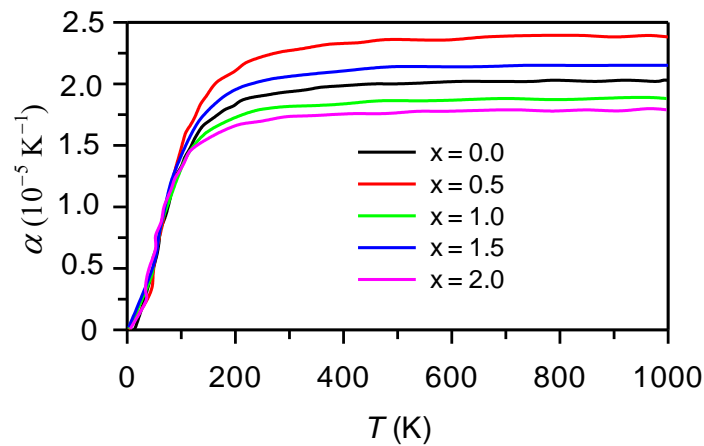

**Fig. 11:** Temperature dependent linear thermal expansion coefficient, $\alpha$, for NaInS$_{2-x}$Se$_x$ (x = 0, 0.5, 1.0, 1.5 and 2.0) solid solutions.

Therefore, it can be said that the low minimum thermal conductivity, comparatively low Debye temperature, moderate values of linear thermal expansion coefficient and damage tolerant behavior clearly indicate NaInS$_{1.5}$Se$_{0.5}$ and NaInS$_{0.5}$Se$_{1.5}$ compounds can be used as promising TBC on other materials for high temperature applications to increase the efficiency of the devices. These solid solutions might even be potential candidates as thermal insulation materials.

5. Conclusions

In summary, the method of first-principles via density functional theory has been employed to investigate the structural and elastic and thermal properties of NaInS$_{2-x}$Se$_x$ (x = 0, 0.5, 1.0, 1.5 and 2.0) solid solutions. It is found that the values of lattice parameters increase with increasing level of Se contents. Cauchy pressure and Poisson's ratio values confirm that the

NaInS$_{1.5}$Se$_{0.5}$ and NaInS$_{0.5}$Se$_{1.5}$ compounds possess significant ionic bonding whereas NaInS$_{2-x}$Se$_x$ (x = 0, 1.0 and 2.0) compounds have mixed (ionic and covalent) bonding character. NaInSSe compound is singled out for its high values of elastic moduli ($Y$, $G$ and $B$) and shows the stiffest behavior, maximum resistant to shear deformation and volume deformation, compared to the other compositions considered here. Ductile behavior is predicted for the compositions x = 0.5 and 1.5 and brittleness is expected for x = 0, 1 and 2 based on the Pugh ratio, $G/B$. Hardness values range from 2.02 to 5.15 GPa, which suggest that all the titled compounds are soft in nature and easily machinable by the conventional cutting tools. The machinability indices can be placed in the sequence: NaInS$_{1.5}$Se$_{0.5}$ > NaInS$_{0.5}$Se$_{1.5}$ > NaInS$_2$ > NaInSe$_2$ > NaInSSe. Elastically anisotropic nature of the solid solutions is confirmed from different anisotropic indices. In particular, the degree of anisotropy in $Y$, $G$ and $v$ increases in the following order: $Y$, $G$, $v$ (NaInS$_{1.5}$Se$_{0.5}$) > $Y$, $G$, $v$ (NaInS$_{0.5}$Se$_{1.5}$) > $Y$, $G$, $v$ (NaInS$_2$) > $Y$, $G$, $v$ (NaInSe$_2$) > $Y$, $G$, $v$ (NaInSSe). The lattice thermal conductivities for NaInS$_{2-x}$Se$_x$ (x = 0, 0.5, 1.0, 1.5 and 2.0) compounds decreased with temperature as 1303.7/T, 624.5/T, 1145/T, 518.9/T and 1086.1/T, respectively. The results of Mulliken bond population analysis suggest that the In-S/Se atoms possess covalent bonds while the Na-S/Se atoms show ionic bonding. The low values of $K_{min}$ originated from the weaker Na-S/Se bonds and comparatively higher mean atomic weight ($M/n$). Some variations of Debye temperature, $\Theta_D$ with composition is due to the variation of polycrystalline elastic moduli, especially the bulk and shear moduli. The values of linear thermal expansion coefficient, ($\alpha$) for NaInS$_{2-x}$Se$_x$ (x = 0, 0.5, 1.0, 1.5 and 2.0) solid solutions are 1.90 × 10$^{-5}$, 2.30 × 10$^{-5}$, 1.80 × 10$^{-5}$, 2.05 × 10$^{-5}$ and 1.70 × 10$^{-5}$ K$^{-1}$ at room temperature, respectively. Favourable thermal ($\Theta_D$, $K_{min}$, $\alpha$ and $\gamma$) and mechanical (ductility and machinability) properties predict that the NaInS$_{1.5}$Se$_{0.5}$ and NaInS$_{0.5}$Se$_{1.5}$ compositions are potential candidate as thermal barrier coating and thermal insulation materials for high temperature device applications. We hope that the present research work will be used as a useful reference for future work.

Acknowledgements

Authors are grateful to the Department of Physics, Chittagong University of Engineering & Technology (CUET), Chattogram-4349, Bangladesh, for providing the financial (Computer) support for this work.

Data availability

The datasets generated during the current study are available from the corresponding author on a reasonable request.